\documentclass[british,english]{article}
\usepackage[T1]{fontenc}
\usepackage[latin9]{inputenc}
\usepackage{geometry}
\usepackage{bm}
\usepackage{amsmath}
\usepackage{amsthm}
\usepackage{amssymb}
\usepackage{stmaryrd}
\usepackage{graphicx}
\usepackage[authoryear]{natbib}

\makeatletter

\providecommand{\tabularnewline}{\\}

\newcommand{\lyxaddress}[1]{
\par {\raggedright #1
\vspace{1.4em}
\noindent\par}
}

\usepackage[hyphens]{url}

\makeatother

\usepackage{babel}
\begin{document}

\title{The fully-visible Boltzmann machine and the Senate of the 45th Australian
Parliament in 2016}

\author{Jessica J. Bagnall$^{*1}$, Andrew T. Jones$^{2}$, \\
Natalie Karavarsamis$^{1}$, and Hien D. Nguyen$^{1}$}
\maketitle

\lyxaddress{$^{*}$Corresponding author email: j.bagnall@latrobe.edu.au. $^{1}$Department
of Mathematics and Statistics, La Trobe University, Bundoora Melbourne,
3086 Australia. $^{2}$School of Mathematics and Physics, University
of Queensland, St. Lucia Brisbane, 4072 Australia.}
\begin{abstract}
After the 2016 double dissolution election, the 45th Australian Parliament
was formed. At the time of its swearing in, the Senate of the 45th
Australian Parliament consisted of nine political parties, the largest
number in the history of the Australian Parliament. Due to the breadth
of the political spectrum that the Senate represented, the situation
presented an interesting opportunity for the study of political interactions
in the Australian context. Using publicly available Senate voting
data in 2016, we quantitatively analyzed two aspects of the Senate.
Firstly, we analyzed the degree to which each of the non-government
parties of the Senate are pro- or anti-government. Secondly, we analyzed
the degree to which the votes of each of the non-government Senate
parties are in concordance or discordance with one another. We utilized
the fully-visible Boltzmann machine (FVBM) model in order to conduct
these analyses. The FVBM is an artificial neural network that can
be viewed as a multivariate generalization of the Bernoulli distribution.
Via a maximum pseudolikelihood estimation approach, we conducted parameter
estimation and constructed hypotheses test that revealed the interaction
structures within the Australian Senate. The conclusions that we drew
are well-supported by external sources of information.
\end{abstract}

\section{Introduction}

At the federal level, or Commonwealth level, the Australian parliamentary
government system is composed of two houses, the House of Representatives,
and the Senate (cf.~\citealp{Weller2003}). The House of Representatives
is composed of members who represent, and are elected by a nearly
equal numbers of voters (currently approximately 150,000 individuals)
from the public. We make a note that the electorates of Tasmania are
notably smaller than the other states (currently approximately 100,000
individuals).

Furthermore, members of the House of Representatives are tasked with
the role of debating legislation and government policy, raising matters
of concern, and importantly making laws via the introduction of bills,
or in the language of the Constitution, proposed laws. Currently,
the House of Representatives consists of 150 members of parliament.
We refer the interested reader to \citet{Wright2012} regarding information
about the House of Representatives of Australia.

In contrast to the House of Representatives, the Senate consists of
76 senators, who do not equally represent the population, but the
states and territories, instead. That is, each of the six states (i.e.~New
South Wales, Queensland, South Australia, Tasmania, Victoria, and
Western Australia) are equally represented by 12 senators, and each
of the two territories (i.e the Australian Capital Territory, and
the Northern Territory) are represented by two senators (cf.~\citealp[Ch. 1]{Evans2016}). 

For any bill to pass to law, both the House of Representatives and
the Senate must assent to the bill, requiring a simple majority vote
in each house. As such, the primary function of the senate is to represent
the people of each of the states and territories equally. It acts
as a balance of power between the states with large populations and
the states with smaller populations, who are not able to secure numbers
in the House of Representatives. Although the Senate does not have
the complete suite of powers to introduce bills, when compared to
the House of Representatives (e.g.~the Senate cannot introduce taxing
bills, appropriation bills, or amend such bills), the Senate does
have the power to reject any bills that are introduced by the House
of Representative, and request the amendment of introduced bills.
More information regarding the Australian Senate and its powers can
be found in \citet{Evans2016}.

In 2016, following a double dissolution election (cf.~\citet{Gauja2018};
\citealp{Corcoran2010} define a double dissolution election as one
where all seats of the House of Representatives and the Senate are
simultaneously contested), the Senate of the 45th Australian Parliament
was formed. In normal election cycles of three years, only half of
the senators from each state face reelection, since each senator is
elected to a six year term (cf.~\citealp[Ch. 1]{Evans2016}). The
territorial senators are elected to three year terms at all elections,
and thus face no special consequence from the special election. The
double dissolution is therefore particularly interesting, as the elected
state senators all face reelection simultaneously, and thus the composition
of the seated Senate is more volatile than after a normal election.
The elected Senate consisted of senators from nine separate political
parties, which is the largest number of political parties in the history
of the Senate (cf.~\citealp[Tab. 2]{Evans2016}). At the time of
its swearing in, the Senate of the 45th Australian Parliament consisted
of 30 senators from the governing Liberal National Party (LNP), 26
senators from the opposition Australian Labour Party (ALP), 9 senators
from the Australian Greens (AG), 4 senators from Pauline Hanson's
One Nation (PHON), 3 senators from the Nick Xenophon Team (NXT), 1
Liberal Democratic Party (LDP) senator, 1 Family First Party (FFP)
senator, 1 Jacqui Lambie Network (JLN) senator, and 1 Derryn Hinch
Justice Party (DHJP) senator. 

When Senators vote on an item of legislation or a motion, the event
is known as a division (cf.~\citealp{Corcoran2010}). Divisions data,
at a party level, are recorded on the official Australian Parliament
website at the URL: \url{www.aph.gov.au/Parliamentary_Business/Statistics/Senate_StatsNet/General/divisions}. 

In this article, we analyze the data from the first sitting of the
Senate of the 45th Australian Parliament, until the final sitting
of the year 2016. The first division during this period was conducted
on the 31st of August 2016, and the last division was performed on
the 1st of December 2016. In total, 147 divisions were performed during
this period. We have chosen this period due to the particular stability
of the Senate during the time. After this period, a number of events,
including the so-called dual citizenship crisis (see, e.g.~\citealp{Begg2017}
and \citealp{Hobbs2018}) resulted in a high number of turnover and
changes in the Senate composition.

We seek to quantitatively analyze two aspects of the Senate. Firstly,
we analyze the degree to which each of the non-government parties
of the Senate are pro- or anti-government. Secondly, we analyze the
degree to which the votes of each of the non-government Senate parties
are in concordance or discordance with one another. In order to answer
both questions we seek to analyze the division data using a fully-visible
Boltzmann machine (FVBM).

The Boltzmann machine (BM) is a parametric generative graphical probabilistic
artificial neural network (ANN) that was introduced in the seminal
paper of \citet{Ackley1985}. The BM is a deep ANN with a latent variable
structure that is capable of universal representation of the probability
mass function (PMF) of multivariate binary random variables of any
fixed dimension $d\in\mathbb{N}$ (cf,~\citealp{LeRoux2008}).

The FVBM, introduced in \citet{Hyvarinen2006}, is a simplification
of the BM, which does not contain a hidden variable structure. It
is a type of log-linear graphical generative probabilistic model that
is comparable to the models that are studied in \citet{Lauritzen1996}.
It can also be considered as a multivariate generalization of the
Bernoulli random variable and is equivalent to the logistic multivariate
binary model proposed by \citet{Cox1972}. Our use of the FVBM is
motivated by that of \citet{Desmarais2010} who construct FVBMs for
the analysis of relationships between judges in the United States
Supreme Court, using data from 178 cases in the period between 2007
and 2008.

In \citet{Hyvarinen2006}, it was proved that an unknown parameter
vector that determines the data generating process (DGP) of data arising
from an FVBM can be estimated consistently via maximum pseudolikelihood
estimation (MPLE) (also known as maximum composite likelihood estimation;
see \citealp{Lindsay1988} and \citealp{Arnold1991}). An alternative
proof of consistency can be found in \citet{Nguyen2018b}. 

Following from the work of \citet{Hyvarinen2006}, \citet{Nguyen2016}
proved that the consistent maximum pseudolikelihood estimator (MPLE)
of an FVBM is also asymptotically normal. Furthermore, in \citet{NguyenWood2016},
it was demonstrated that the MPLE for an FVBM could be efficiently
computed from data via an iterative block-successive lower bound maximization
algorithm of the kind that is described in \citet{Razaviyayn2013}.
It was proved that the computational algorithm monotonically increases
the pseudolikelihood objective in each iteration, and that the limit
point of the algorithm is a global maximum of the objective function.
In this paper, we utilize the \emph{R} package \emph{BoltzMM }\citep{Jones2018},
which implements the methods and algorithms that are described in
\citet{Nguyen2016} and \citet{NguyenWood2016}, for the analysis
of the our Senate data.

The remainder of the article proceeds as follows. In Section 2, we
describe the FVBM, the MPLE estimator, and the inferential tools that
we will use. In Section 3, we describe our Senate data in detail,
and we discuss the details of our data processing protocol. In Section
4, we present the result of an analysis of the Senate data via an
FVBM and we discuss the results of the analysis. In Section 5, we
draw some conclusions regarding our work. Auxiliary and technical
results are presented in the Appendix.

\section{The fully-visible Boltzmann machine}

Let $\bm{X}^{\top}=\left(X_{1},\dots,X_{d}\right)$ be a vector of
spin binary random variables (i.e.~$X_{j}\in\left\{ -1,+1\right\} $;
$j\in\left[d\right]=\left\{ 1,\dots,d\right\} $). Here, $\left(\cdot\right)^{\top}$
is the transposition operator. We say that the DGP of $\bm{X}$ is
a FVBM if it can be characterized by the PMF
\begin{align}
f\left(\bm{x};\bm{\theta}\right) & =\mathbb{P}\left(\bm{X}=\bm{x}\right)\nonumber \\
 & =\frac{\exp\left(\frac{1}{2}\bm{x}^{\top}\bm{M}\bm{x}+\bm{x}^{\top}\bm{b}\right)}{z\left(\bm{\theta}\right)}\text{,}\label{eq: PMF}
\end{align}
where
\[
z\left(\bm{\theta}\right)=\sum_{\bm{\xi}\in\left\{ -1,1\right\} ^{d}}\exp\left(\frac{1}{2}\bm{\xi}^{\top}\bm{M}\bm{\xi}+\bm{\xi}^{\top}\bm{b}\right)\text{,}
\]
$\bm{M}\in\mathbb{R}^{d\times d}$ is a symmetric matrix with zeros
along the diagonal, and $\bm{b}\in\mathbb{R}^{d}$. We call $z\left(\bm{\theta}\right)$
the normalization constant, and we put the unique elements of $\bm{M}$
and $\bm{b}$ into the $\left[n\left(n-1\right)/2+n\right]$-dimensional
parameter vector $\bm{\theta}$.

The bias vector $\bm{b}$ can be written as:
\[
\bm{b}^{\top}=\left(b_{1},\dots,b_{d}\right)\text{,}
\]
where increasing the value of $b_{j}\in\mathbb{R}$ ($j\in\left[d\right]$)
increases the probability $\mathbb{P}\left(X_{j}=+1\right)$, ceteris
paribus. Similarly, decreasing the value of $b_{j}$ increases the
probability $\mathbb{P}\left(X_{j}=-1\right)$, ceteris paribus.

The interaction matrix $\bm{M}$ can be written as:
\[
\bm{M}=\left[\begin{array}{ccccc}
0 & m_{12} & m_{13} & \dots & m_{1d}\\
m_{12} & 0 & m_{23} & \dots & m_{2d}\\
m_{13} & m_{23} & 0 & \ddots & \vdots\\
\vdots & \vdots & \ddots & \ddots & m_{d-1,d}\\
m_{1d} & m_{2d} & \cdots & m_{d-1,d} & 0
\end{array}\right]\text{,}
\]
where increasing the value of $m_{jk}\in\mathbb{R}$ ($j,k\in\left[d\right]$;
$j<k$) increases the probability $\mathbb{P}\left(X_{j}=X_{k}\right)$,
ceteris paribus. Similarly, decreases in the value of $m_{jk}$ increases
the probability $\mathbb{P}\left(X_{j}\ne X_{k}\right)$, ceteris
paribus. Thus, the vector $\bm{b}$ controls the marginal probabilities
of each element of $\bm{X}$, whereas the matrix $\bm{M}$ controls
the correlations between the elements of $\bm{X}$.

\subsection{Estimation of the FVBM parameter vector}

Suppose that we observe $n$ independent and identical replicates
of $\bm{X}$, say $\bm{X}_{1},\dots,\bm{X}_{n}$, which arise from
a DGP that can be characterized by an FVBM with unknown parameter
vector $\bm{\theta}_{0}$. Since the FVBM is a probabilistic model
with PMF (\ref{eq: PMF}), we can construct the log-likelihood function
\begin{align*}
l_{n}\left(\bm{\theta}\right) & =\sum_{i=1}^{n}\log f\left(\bm{X}_{i};\bm{\theta}\right)\\
 & =\frac{1}{2}\sum_{i=1}^{n}\bm{X}_{i}^{\top}\bm{M}\bm{X}_{i}+\sum_{i=1}^{n}\bm{X}_{i}^{\top}\bm{b}-n\log z\left(\bm{\theta}\right)\text{,}
\end{align*}
and estimate $\bm{\theta}_{0}$ via the maximum likelihood estimator
(MLE)
\[
\hat{\bm{\theta}}_{n}=\arg\max_{\bm{\theta}}\text{ }l_{n}\left(\bm{\theta}\right)\text{.}
\]

Unfortunately, for all but the smallest values of $d$, the log-likelihood
function is prohibitively computationally expensive to work with,
since the number of terms in $z\left(\bm{\theta}\right)$ grows exponentially
with $d$. Due to the computational costs of obtaining the MLE, it
is more common to consider the MPLE as an estimator of $\bm{\theta}_{0}$
in the FVBM context, instead (see, e.g.~\citealp{Hyvarinen2006},
\citealp{Desmarais2010}, \citealp{NguyenWood2016}, and \citealp{Nguyen2016}).

The so-called log-pseudolikelihood function, as used by the references
above, can be written as
\begin{equation}
p_{n}\left(\bm{\theta}\right)=\sum_{i=1}^{n}\sum_{j=1}^{d}\log f\left(X_{ij}|\bm{X}_{i\left(j\right)};\bm{\theta}\right)\text{,}\label{eq: PL fun}
\end{equation}
where $\bm{X}_{\left(j\right)}^{\top}=\left(X_{1},\dots,X_{j-1},X_{j+1},\dots,X_{d}\right)$
and 
\[
f\left(x_{j}|\bm{x}_{\left(j\right)};\bm{\theta}\right)=\frac{\exp\left(x_{j}\bm{m}_{j}^{\top}\bm{x}+b_{j}x_{j}\right)}{\exp\left(\bm{m}_{j}^{\top}\bm{x}+b_{j}\right)+\exp\left(-\bm{m}_{j}^{\top}\bm{x}-b_{j}\right)}\text{.}
\]
Here, $\bm{m}_{j}$ is the $j\text{th}$ column of $\bm{M}$, and
we call $\prod_{j=1}^{d}f\left(X_{j}|\bm{X}_{\left(j\right)};\bm{\theta}\right)$
the individual pseudolikelihood of observation $\bm{X}$. Note that
(\ref{eq: PL fun}) no longer depends on the normalization constant
$z\left(\bm{\theta}\right)$. Like the MLE, the MPLE can be defined
as the maximizer of (\ref{eq: PL fun}). That is, we can write the
MPLE as
\begin{equation}
\tilde{\bm{\theta}}_{n}=\arg\max_{\bm{\theta}}\text{ }p_{n}\left(\bm{\theta}\right)\text{.}\label{eq: MPLE}
\end{equation}

\subsection{\label{subsec: Asymptotic normality}Properties of the MPLE}

The MPLE has a number of desirable asymptotic properties. Firstly,
via the results of \citet{Hyvarinen2006} and \citet{Nguyen2018b},
the MPLE is proved to be a consistent estimator of $\bm{\theta}_{0}$
in the case when the FVBM is a well-specified model for the DGP of
the data, and is proved to be Wald consistent, in the sense of \citet[Sec. 5.2.1]{vanderVaart1998},
when the FVBM is a misspecified model, respectively.

Furthermore, in either case (defining $\bm{\theta}_{0}$ to be the
true parameter vector in the well-specified case, or the asymptotic
global maximum of the expected log individual pseudolikelihood in
the misspecified case), the main theorem of \citet{Nguyen2016} yields
the asymptotic normality of $n^{1/2}\left(\tilde{\bm{\theta}}_{n}-\bm{\theta}_{0}\right)$.
That is, as $n$ approaches infinity, $n^{1/2}\left(\tilde{\bm{\theta}}_{n}-\bm{\theta}_{0}\right)$
converges in law to a multivariate normal distribution with mean vector
$\mathbf{0}$ (the zero vector) and covariance matrix $\bm{I}_{1}^{-1}\left(\bm{\theta}_{0}\right)\bm{I}_{2}\left(\bm{\theta}_{0}\right)\bm{I}_{1}^{-1}\left(\bm{\theta}_{0}\right)$,
where
\[
\bm{I}_{1}\left(\bm{\theta}\right)=-\sum_{j=1}^{d}\mathbb{E}\left[\frac{\partial^{2}\log f\left(X_{j}|\bm{X}_{\left(j\right)};\bm{\theta}\right)}{\partial\bm{\theta}\partial\bm{\theta}^{\top}}\right]\text{,}
\]
and 
\[
\bm{I}_{2}\left(\bm{\theta}\right)=\sum_{j=1}^{d}\sum_{k=1}^{d}\mathbb{E}\left[\frac{\partial\log f\left(X_{j}|\bm{X}_{\left(j\right)};\bm{\theta}\right)}{\partial\bm{\theta}}\cdot\frac{\partial\log f\left(X_{j}|\bm{X}_{\left(j\right)};\bm{\theta}\right)}{\partial\bm{\theta}^{\top}}\right]\text{.}
\]

Using the estimator $\tilde{\bm{\theta}}_{n}$, we can consistently
estimate the information matrices $\bm{I}_{1}\left(\bm{\theta}_{0}\right)$
and $\bm{I}_{2}\left(\bm{\theta}_{0}\right)$ via the empirical information
matrices $\tilde{\bm{I}}_{1}\left(\tilde{\bm{\theta}}_{n}\right)$
and $\tilde{\bm{I}}_{2}\left(\tilde{\bm{\theta}}_{n}\right)$, respectively
(cf.~\citealp[Sec. 7.8.3]{Boos2013}), where
\[
\tilde{\bm{I}}_{1}\left(\bm{\theta}\right)=-\frac{1}{n}\sum_{i=1}^{n}\sum_{j=1}^{d}\frac{\partial\log f\left(X_{ij}|\bm{X}_{i\left(j\right)};\bm{\theta}\right)}{\partial\bm{\theta}\partial\bm{\theta}^{\top}}
\]
and
\[
\tilde{\bm{I}}_{2}\left(\bm{\theta}\right)=\frac{1}{n}\sum_{i=1}^{n}\sum_{j=1}^{d}\sum_{k=1}^{d}\frac{\partial\log f\left(X_{ij}|\bm{X}_{i\left(j\right)};\bm{\theta}\right)}{\partial\bm{\theta}}\cdot\frac{\partial\log f\left(X_{ij}|\bm{X}_{i\left(j\right)};\bm{\theta}\right)}{\partial\bm{\theta}^{\top}}\text{.}
\]

\subsection{\label{subsec: BSLB}The block-successive lower bound maximization
algorithm}

Using the framework of block-successive lower bound maximization algorithms
(also known as block minormization-maximization algorithms; cf.~\citealp{Hunter2004}
and \citealp{Nguyen2017}), \citet{NguyenWood2016} proposed the following
iterative algorithm for the computation of (\ref{eq: MPLE}), upon
observing a realization $\bm{x}_{1},\dots,\bm{x}_{n}$ of $\bm{X}_{1},\dots,\bm{X}_{n}$.

Denote an initial estimate of the MPLE $\bm{\theta}^{\left(0\right)}$
and write the $r\text{th}$ iterate of the algorithm as $\bm{\theta}^{\left(r\right)}$
(containing the components $\bm{M}^{\left(r\right)}$ and $\bm{b}^{\left(r\right)}$).
At the $r\text{th}$ iteration, in the order $j=1,2,\dots,d$, we
compute

\begin{equation}
b_{j}^{\left(r\right)}=\frac{1}{n}\sum_{i=1}^{n}\left[x_{ij}-\tanh\left(\bm{m}_{j}^{\left(r-1\right)\top}\bm{x}_{i}+b_{j}^{\left(r-1\right)}\right)\right]+b_{j}^{\left(r-1\right)}\text{.}\label{eq: bup}
\end{equation}

Let $m_{jk}^{\left(r\right)}$ be the $j\text{th}$ row and $k\text{th}$
column element of $\bm{M}^{\left(r\right)}$, and let $\bm{M}_{\left[uv\right]}^{\left(s\right)}$
be a symmetric matrix with zeros along the diagonal and elements
\[
m_{\left[uv\right]jk}^{\left(r\right)}=\begin{cases}
m_{jk}^{\left(r+1\right)}\text{,} & \text{if }j<u\text{, or }j=u\text{ and }k<v\text{,}\\
m_{jk}^{\left(r\right)}\text{,} & \text{otherwise.}
\end{cases}
\]
Then, in the lexicographical order
\[
\left(i,j\right)=\left(1,2\right),\dots,\left(1,d\right),\left(2,3\right),\left(2,4\right),\dots,\left(d-1,d\right)\text{,}
\]
we compute
\begin{eqnarray}
m_{jk}^{\left(r\right)} & = & \frac{1}{2n}\sum_{i=1}^{n}\left[2x_{ij}x_{ik}-x_{ik}\tanh\left(\bm{m}_{\left[jk\right]j}^{\left(r-1\right)\top}\bm{x}_{i}+b_{j}^{\left(r\right)}\right)-x_{ij}\tanh\left(\bm{m}_{\left[jk\right]k}^{\left(r-1\right)\top}\bm{x}_{i}+b_{k}^{\left(r\right)}\right)\right]\nonumber \\
 &  & +m_{jk}^{\left(r-1\right)}\text{,}\label{eq: mup}
\end{eqnarray}
where $\bm{m}_{\left[uv\right]j}^{\left(r\right)}$ is the $j\text{th}$
column of the matrix $\bm{M}_{\left[uv\right]}^{\left(r\right)}$.
The iterations (\ref{eq: bup}) and (\ref{eq: mup}) are repeated
until some convergence criterion is meet (e.g., until $r=R$, where
$R$ is some large number). The final iterate is then used as the
MPLE estimator $\tilde{\bm{\theta}}_{n}$.

In \citet{NguyenWood2016}, it is proved that the algorithm defined
by (\ref{eq: bup}) and (\ref{eq: mup}) is monotonic, in the sense
that $p_{n}\left(\bm{\theta}^{\left(r\right)}\right)\le p_{n}\left(\bm{\theta}^{\left(r+1\right)}\right)$,
for each $r\in\mathbb{N}$. That is, the algorithm increases the log-pseudolikelihood
objective in each iteration. Furthermore, it is also proved that for
any starting value $\bm{\theta}^{\left(0\right)}$, the limit point
of the algorithm $\bm{\theta}^{\left(\infty\right)}=\lim_{r\rightarrow\infty}\bm{\theta}^{\left(r\right)}$
is the global maximizer of the log-pseudolikelihood function $p_{n}\left(\bm{\theta}\right)$.
Thus, the algorithm is globally convergent to the global maximizer
of the objective, regardless of how it is initialized. The two results
guarantee the stability and correctness of the described algorithm.

\subsection{\label{subsec:Hypothesis-testing}Hypothesis testing}

Using the Wald statistics construction of \citet[Sec. 9.3.1]{Molenberghs2005},
we can construct a statistic to test the hypotheses that
\[
\text{H}_{0}:\text{ }\theta_{k}=\theta_{0k}\text{, versus H}_{1}:\text{ }\theta_{k}\ne\theta_{0k}\text{,}
\]
where $\theta_{k}$ ($\theta_{0k}$) is the $k\text{th}$ element
of $\bm{\theta}$ ($\bm{\theta}_{0}$), for $k\in\left[n\left(n-1\right)/2+n\right]$.
For each $k$, the $z$-score form of the Wald statistic can be given
as
\begin{equation}
Z_{k}=\frac{\hat{\theta}_{nk}-\theta_{0k}}{\sqrt{\hat{\sigma}_{nk}^{2}/n}}\text{,}\label{eq: Z stat}
\end{equation}
where $\hat{\sigma}_{n,k}^{2}$ is the $k\text{th}$ diagonal element
of the matrix $\tilde{\bm{I}}_{1}^{-1}\left(\tilde{\bm{\theta}}_{n}\right)\tilde{\bm{I}}_{2}\left(\tilde{\bm{\theta}}_{n}\right)\tilde{\bm{I}}_{1}^{-1}\left(\tilde{\bm{\theta}}_{n}\right)$.
Under the null hypothesis $\text{H}_{0}$, $Z_{k}$ is asymptotically
standard normal. This asymptotic result can be used as an approximation
in order to construct finite-sample hypothesis tests.

\section{\label{sec: Data set}The Senate data}

The data that we study are taken directly from \url{www.aph.gov.au/Parliamentary_Business/Statistics/Senate_StatsNet/General/divisions}.
In particular, we study the divisions data taken from the first sitting
of the Senate of the 45th Australian Parliament, until the final sitting
during the 2016 calendar year.

The data contains $n=147$ rows and each of the columns of the data
contains an indicator of how the particular party (i.e.~LNP, ALP,
AG, NXT, PHON, LDP, FFP, JLN, and DHJP) voted on each of the $n$
items of legislation. The indicators were ``Yes'' (indicating a
vote in favor of the legislation), ``No'' (indicating a vote against
the legislation), ``Split'' (indicating that the members of the
party did not all vote in the same direction), and ``-/{[}blank{]}''
(indicating no vote cast). 

\subsection{\label{subsec: Data-processing}Data processing}

Our first preprocessing step is to investigate and decide upon whether
the ``Split'' indicator should be relabeled as ``Yes'' or ``No''.
Upon investigation, we notice that all of the ``Split'' entries
arose from the same party: PHON. Although tedious, it was possible
to ascertain the manner in which the different members of PHON voted
in each of the 12 ``Split'' votes. These data were obtained from
the Senate Journal, which are textual documents that contain the untabulated
data regarding votes on Senate questions at the individual level,
within their contents. The Senate Journal documents can be obtained
at \url{parlinfo.aph.gov.au}. Table \ref{tab: PHON split} presents
the voting pattern of the four members of PHON (Senators Burston,
Culleton, Hanson, and Roberts) on the 12 ``Split'' entries.

\begin{table}
\caption{\label{tab: PHON split}Voting patterns for each of the PHON senators
on the 12 ``Split'' entries within the divisions data. All dates
are in the year 2016 and the number indicates which division item
on the date that the vote corresponded to.}

\centering{}%
\begin{tabular}{cccccc}
\hline 
Date & Number & Burston & Culleton & Hanson & Roberts\tabularnewline
\hline 
13/9 & 2 & No & Yes & No & No\tabularnewline
14/9 & 4 & Yes & No & Yes & Yes\tabularnewline
23/11 & 1 & No & No & Yes & Yes\tabularnewline
29/11 & 2 & No & Yes & No & No\tabularnewline
29/11 & 3 & No & Yes & No & No\tabularnewline
29/11 & 4 & No & Yes & No & No\tabularnewline
30/11 & 7 & No & Yes & No & No\tabularnewline
30/11 & 8 & No & Yes & No & No\tabularnewline
30/11 & 11 & No & Yes & No & No\tabularnewline
01/12 & 4 & No & Yes & No & -\tabularnewline
01/12 & 10 & No & Yes & No & No\tabularnewline
01/12 & 25 & No & Yes & No & No\tabularnewline
\hline 
\end{tabular}
\end{table}

Upon inspecting the details of Table \ref{tab: PHON split}, we observe
that Senator Culleton went against the party leader (Senator Hanson)
on every one of the 12 occasions. The only other person to vote in
the opposite direction to Senator Hanson was Senator Burston, on 23
November. The table also shows that Senator Roberts missed division
number 4 on 1 December.

On 18 December 2016, Senator Culleton resigned from PHON and sat as
an independent Senator. Furthermore, on 23 December 2016, the Federal
Court of Australia found that Senator Culleton was bankrupt, which
raised questions regarding his eligibility to remain on the Senate
(cf.~\citealp{Phillips2017}). It was not until February 2017 that
Senator Culleton was disqualified from the Senate by the High Court
of Australia (cf.~\citealp{Stubbs2017}). The disqualification was
largely due to the fact that the Senator was convicted of a criminal
offense, prior to his election, and thus was not eligible to stand
for the Senate during the 2016 election \citep{Remeikis2017}.

We can conclude that Senator Culleton was influential, as an independent
entity, during the year of 2016. We choose to handle the ``Split''
entries in the following way: firstly, Senator Culleton will be treated
as a separate entity to the rest of PHON, and secondly, the majority
vote is taken as the indicator for the rest of the PHON entries. We
note that in all but one case, the majority vote is also the unanimous
vote.

Next, the indicator ``-/{[}blank{]}'' was taken to be a missing
data entry in the context of our analysis. Taken as such, we found
that the FFP column contained 142 missing data entries. Upon review,
the large number of missing entries in the FFP column was due to various
ineligibility issues regarding the only FFP Senator at the time, Bob
Day (cf.~\citealp{Stubbs2017}). Since the FFP played a minimal role
in the Senate during the investigated period, we eliminated the FFP
column from our data.

The remainder of the data contained a further 54 missing entries out
of a potential $147\times9=1323$ entries. We considered this missingness
be a minor issue (only $4.08\%$) and used 3-nearest neighbor imputation
(cf.~\citealp{Troyanskaya2001}) in order to fill in the missing
entries. The choice of 3-nearest neighbor imputation was made to balance
between precision and generalizability, as suggested in \citet{Beretta2016}.
We conducted the imputation via the function \emph{knn.impute()} within
the \emph{R} package: \emph{bnstruct} \citep{Franzin2017}. See \citet{RCT2016}
regarding \emph{R}. We note that other levels of $k$ for $k$-nearest
neighbor imputation were also assessed, although we found that our
inference was insensitive to the choice of $k$ (see the Appendix
for details).

Next, we construct the binary random variables of interest: the agreement
of each Senate party's vote with the Government's vote on each item
of legislation. For each of the non-government parties (i.e.~not
LNP) we construct new columns of data. The new entries of data can
be given as $x_{ij}$ where $i$ denote the index of the legislation
and $j$ denotes the index of the party. We set $x_{ij}$ to $+1$
if both the LNP and the $j\text{th}$ party voted ``Yes'' or ``No''
on legislation $i$. We set $x_{ij}$ to $-1$ if the LNP voted ``Yes''
and the $j\text{th}$ party voted ``No'', or vice versa, if the
LNP voted ``No'' and the $j\text{th}$ party voted ``Yes''. For
each item of legislation $i\in\left[n\right]$ ($n=147$), the vector
$\bm{x}_{i}^{\top}=\left(x_{i1},\dots,x_{id}\right)$ ($d=8$) summaries
the agreement or disagreement of each of the non-government Senate
parties (excluding the FFP, and including Senator Culleton, which
we code as CULL, as his own entity) with the Government. 

\section{\label{sec: Analysis}Analysis of the Senate data}

The analysis of the processed Senate data was conducted using the
\emph{R }package \emph{BoltzMM}. The main function of the package
is the \emph{fitfvbm} function, which applies the algorithm from Section
\ref{subsec: BSLB} in order to compute the MPLE vector $\tilde{\bm{\theta}}_{n}$.
The functions \emph{fvbmHess}, \emph{fvbmcov}, and \emph{fvbmstderr},
can then be applied to compute the estimated standard errors for each
element of $\tilde{\bm{\theta}}_{n}$, under the normal approximation
afforded by the asymptotic normality of the MPLE (i.e.~the root diagonal
elements of the matrix $n^{-1}\bm{I}_{1}^{-1}\left(\tilde{\bm{\theta}}_{n}\right)\bm{I}_{2}\left(\tilde{\bm{\theta}}_{n}\right)\bm{I}_{1}^{-1}\left(\tilde{\bm{\theta}}_{n}\right)$).

Using the functions from \emph{BoltzMM}, we computed the elements
of MPLE $\tilde{\bm{\theta}}_{n}$ for the Senate data and we estimated
the standard error for each element of $\tilde{\bm{\theta}}_{n}$.
The MPLE and standard error estimates are presented in Tables \ref{tab: Estimates}
and \ref{tab: SErrs}, respectively.

\begin{table}
\caption{\label{tab: Estimates}Sub-table A contains the estimated values of
the bias vector (i.e.~$\tilde{\bm{b}}_{n}$). Sub-table B contains
the estimated values of the unique elements of the interaction matrix
(i.e $\tilde{\bm{M}}_{n}$).}

\centering{}%
\begin{tabular}{ccccccccc}
A &  &  &  &  &  &  &  & \tabularnewline
\hline 
Party & ALP & AG & NXT & PHON & LDP & JLN & DHJP & CULL\tabularnewline
\hline 
Bias & -0.321 & -1.037 & -0.209 & 0.941 & 0.384 & -0.559 & 0.693 & -0.383\tabularnewline
\hline 
 &  &  &  &  &  &  &  & \tabularnewline
B &  &  &  &  &  &  &  & \tabularnewline
\cline{1-8} 
Party & ALP & AG & NXT & PHON & LDP & JLN & DHJP & \tabularnewline
\cline{1-8} 
AG & -0.203 &  &  &  &  &  &  & \tabularnewline
NXT & -0.185 & -0.284 &  &  &  &  &  & \tabularnewline
PHON & -0.370 & 0.147 & 0.371 &  &  &  &  & \tabularnewline
LDP & 0.173 & -0.053 & -0.208 & 0.512 &  &  &  & \tabularnewline
JLN & 0.321 & 0.626 & 0.419 & 0.394 & 0.024 &  &  & \tabularnewline
DHJP & 0.059 & 0.601 & 0.808 & -0.498 & 0.224 & 0.077 &  & \tabularnewline
CULL & 0.042 & -0.710 & -0.146 & 1.287 & 0.116 & 0.397 & 0.801 & \tabularnewline
\cline{1-8} 
\end{tabular}
\end{table}

\begin{table}
\caption{\label{tab: SErrs}Sub-table A contains the estimated standard errors
for the MPLE bias vector. Sub-table B contains the estimated standard
errors of the unique elements of the MPLE interaction matrix.}

\centering{}%
\begin{tabular}{ccccccccc}
A &  &  &  &  &  &  &  & \tabularnewline
\hline 
Party & ALP & AG & NXT & PHON & LDP & JLN & DHJP & CULL\tabularnewline
\hline 
St. Err. & 0.165 & 0.207 & 0.208 & 0.274 & 0.164 & 0.194 & 0.239 & 0.326\tabularnewline
\hline 
 &  &  &  &  &  &  &  & \tabularnewline
B &  &  &  &  &  &  &  & \tabularnewline
\cline{1-8} 
Party & ALP & AG & NXT & PHON & LDP & JLN & DHJP & \tabularnewline
\cline{1-8} 
AG & 0.130 &  &  &  &  &  &  & \tabularnewline
NXT & 0.130 & 0.218 &  &  &  &  &  & \tabularnewline
PHON & 0.212 & 0.240 & 0.268 &  &  &  &  & \tabularnewline
LDP & 0.130 & 0.141 & 0.158 & 0.219 &  &  &  & \tabularnewline
JLN & 0.117 & 0.154 & 0.141 & 0.216 & 0.131 &  &  & \tabularnewline
DHJP & 0.144 & 0.248 & 0.163 & 0.335 & 0.149 & 0.157 &  & \tabularnewline
CULL & 0.212 & 0.274 & 0.232 & 0.260 & 0.211 & 0.178 & 0.337 & \tabularnewline
\cline{1-8} 
\end{tabular}
\end{table}

Via the Wald hypothesis test construction from Section \ref{subsec:Hypothesis-testing},
and using the results from Tables \ref{tab: Estimates} and \ref{tab: SErrs},
we conducted tests of the null hypotheses that $\theta_{k}=0$, versus
the alternative hypothesis that $\theta_{k}\ne0$, for each $k\in\left[n\left(n-1\right)/2+n\right]$,
where each $\theta_{k}$ is an element of the true parameter vector
$\bm{\theta}$, which specifies the DGP of the observations. The test
statistic for each $k$ is computed as per (\ref{eq: Z stat}). The
$p\text{-value}$ of each test, computed using normal approximations
as justified by the results of Section \ref{subsec: Asymptotic normality},
are presented in Table \ref{tab: p-values}.

\begin{table}
\caption{\label{tab: p-values}Sub-table A contains the $p\text{-values}$
for the tests that each bias vector element is equal to zero. Sub-table
B contains the $p\text{-values}$ for the tests that each interaction
matrix element is equal to zero.}

\centering{}{\footnotesize{}}%
\begin{tabular}{ccccccccc}
{\footnotesize{}A} &  &  &  &  &  &  &  & \tabularnewline
\hline 
{\footnotesize{}Party} & {\footnotesize{}ALP} & {\footnotesize{}AG} & {\footnotesize{}NXT} & {\footnotesize{}PHON} & {\footnotesize{}LDP} & {\footnotesize{}JLN} & {\footnotesize{}DHJP} & {\footnotesize{}CULL}\tabularnewline
\hline 
{\footnotesize{}$p\text{-val}$} & {\footnotesize{}5.20E-02} & {\footnotesize{}5.71E-07} & {\footnotesize{}3.13E-01} & {\footnotesize{}5.89E-04} & {\footnotesize{}1.94E-02} & {\footnotesize{}3.92E-03} & {\footnotesize{}3.79E-03} & {\footnotesize{}2.40E-01}\tabularnewline
\hline 
 &  &  &  &  &  &  &  & \tabularnewline
{\footnotesize{}B} &  &  &  &  &  &  &  & \tabularnewline
\cline{1-8} 
{\footnotesize{}Party} & {\footnotesize{}ALP} & {\footnotesize{}AG} & {\footnotesize{}NXT} & {\footnotesize{}PHON} & {\footnotesize{}LDP} & {\footnotesize{}JLN} & {\footnotesize{}DHJP} & \tabularnewline
\cline{1-8} 
{\footnotesize{}AG} & {\footnotesize{}1.16E-01} &  &  &  &  &  &  & \tabularnewline
{\footnotesize{}NXT} & {\footnotesize{}1.56E-01} & {\footnotesize{}1.92E-01} &  &  &  &  &  & \tabularnewline
{\footnotesize{}PHON} & {\footnotesize{}8.13E-02} & {\footnotesize{}5.39E-01} & {\footnotesize{}1.66E-01} &  &  &  &  & \tabularnewline
{\footnotesize{}LDP} & {\footnotesize{}1.84E-01} & {\footnotesize{}7.09E-01} & {\footnotesize{}1.88E-01} & {\footnotesize{}1.91E-02} &  &  &  & \tabularnewline
{\footnotesize{}JLN} & {\footnotesize{}5.87E-03} & {\footnotesize{}5.01E-05} & {\footnotesize{}2.90E-03} & {\footnotesize{}6.75E-02} & {\footnotesize{}8.56E-01} &  &  & \tabularnewline
{\footnotesize{}DHJP} & {\footnotesize{}6.85E-01} & {\footnotesize{}1.53E-02} & {\footnotesize{}7.24E-07} & {\footnotesize{}1.37E-01} & {\footnotesize{}1.34E-01} & {\footnotesize{}6.21E-01} &  & \tabularnewline
{\footnotesize{}CULL} & {\footnotesize{}8.42E-01} & {\footnotesize{}9.55E-03} & {\footnotesize{}5.29E-01} & {\footnotesize{}7.58E-07} & {\footnotesize{}5.83E-01} & {\footnotesize{}2.57E-02} & {\footnotesize{}1.74E-02} & \tabularnewline
\cline{1-8} 
\end{tabular}{\footnotesize \par}
\end{table}

The function \emph{allpfvbm }allows for the calculation of the probability
of all possible outcomes, under the fitted FVBM. Summing over the
probabilities for each party, we can obtain the marginal probabilities
of each party voting in correspondence with the government on each
Senate question. These model-estimated probabilities are reported
in Table \ref{tab: Probabilities}. We also report, for comparison,
the empirical proportions computed as the proportion $\bar{p}_{j}=n^{-1}\sum_{i=1}^{n}\left\llbracket x_{ij}=+1\right\rrbracket $
for each party $j$, where $\left\llbracket \cdot\right\rrbracket $
is the Iverson bracket notation for the indicator function \citep[Ch. 2]{Graham1989}.

\begin{table}
\caption{\label{tab: Probabilities}The first row of the table contains model-based
probability estimates of the marginal probabilities of agreement $\mathbb{P}\left(X_{j}=+1\right)$,
for each Senate party $j$, computed using the function \emph{allpfvbm}.
The second row contains the corresponding empirical proportions.}

\centering{}%
\begin{tabular}{ccccccccc}
\hline 
Party & ALP & AG & NXT & PHON & LDP & JLN & DHJP & CULL\tabularnewline
\hline 
Model & 0.330 & 0.122 & 0.581 & 0.815 & 0.784 & 0.342 & 0.649 & 0.761\tabularnewline
Empirical & 0.333 & 0.129 & 0.592 & 0.810 & 0.782 & 0.354 & 0.660 & 0.755\tabularnewline
\hline 
\end{tabular}
\end{table}

Upon observation of Table \ref{tab: Probabilities}, we firstly notice
that the probabilities obtained under the fitted FVBM model closely
resemble those that are obtained via the computation of empirical
proportions. We note that the smallest standard error (calculated
using the usual expression $\left[\bar{p}_{j}\left(1-\bar{p}_{j}\right)/n\right]^{1/2}$,
using the proportion for the party AG) is $0.028$, and thus the smallest
approximately-normal $95\%$ confidence interval margin of error over
all of the proportions is $2\times0.028=0.056$. We see that the model-estimated
probabilities all fall comfortably within the $95\%$ confidence interval
of the empirical proportions, and thus the two sets of estimates can
be seen to be sufficiently close.

Secondly, we notice that all but two of the model-estimated probabilities
(or empirical proportions) correspond well with the sign of the estimated
bias parameter elements (from Table \ref{tab: Estimates}A). That
is, ALP, AG, and JLN each have negative bias estimates and each vote
in agreement with the government less than $50\%$ of the time. We
also observe that PHON, LDP, and DHJP all have positive estimated
biases and all vote in agreement with the government greater than
$50\%$ of the time.

The two situations where the biases and marginal probabilities are
in different direction are for NXT and CULL. Both NXT and CULL have
negative biases but vote in agreement with the government more than
$50\%$ of the time. In both cases, the differences in the direction
can be attributed to the strong interactions positive interaction
between the two parties under discussion and other parties that are
biased towards agreement with the government (namely PHON and DHJP).

From Table \ref{tab: Estimates}B, we observe that the interaction
coefficient between NXT and DHJP is the second highest estimated value,
at $0.808$. Using the function \emph{allpfvbm}, we can obtain the
joint PMF between the two parties. The estimated joint PMF indicates
that both parties vote in agreement with the government with probability
$0.528$ and both parties vote against the government with probability
$0.298$. Furthermore, NXT votes with the government whilst DHJP votes
against the government with probability $0.053$, and NXT votes against
the government whilst DHJP votes with the government with probability
$0.121$. From these values, we can see that the two parties vote
in concordance with one another with at an estimated rate of $82.6\%$
of the time. Thus, it is not surprising that although NXT has a slightly
negative bias, that the strong relationship with DHJP influences NXT
to vote in agreement with the government at a rate that is higher
than $50\%$.

Again, from Table \ref{tab: Estimates}B, we observe that the interaction
coefficient between CULL and PHON is the highest estimated value,
at $1.287$. This is unsurprising, since we know from Table \ref{tab: PHON split}
that CULL and PHON agree on all but 12 Senate items. Using the function
\emph{allpfvbm}, we can obtain the joint PMF between the two parties.
The estimated joint PMF indicates that both parties vote in agreement
with the government with probability $0.746$ and both parties vote
against the government with probability $0.170$. Furthermore, CULL
votes with the government whilst PHON votes against the government
with probability $0.015$, and CULL votes against the government whilst
PHON votes with the government with probability $0.069$. The joint
PMF thus indicates that the two parties vote in concordance with one
another at an estimated rate of $91.6\%$. This explains why CULL
has an agreement rate with the government that is greater than $50\%$,
even while the estimated bias is negative.

From the results that were presented above, we can remark that one
cannot therefore interpret the bias estimates $\tilde{\bm{b}}_{n}$
and the empirical proportions $\bar{p}_{j}$ (and model-estimated
probabilities) in the same manner. The biases provide an interpretation
of the behavior of each party when interactions between the parties
are simultaneously being accounted for. Alternatively, the empirical
proportions and model-estimated probabilities can provide an interpretation
of the behavior of each party, marginally, without accounting for
the their interaction with the other Senate entities.

\begin{figure}
\begin{centering}
\includegraphics[width=12cm]{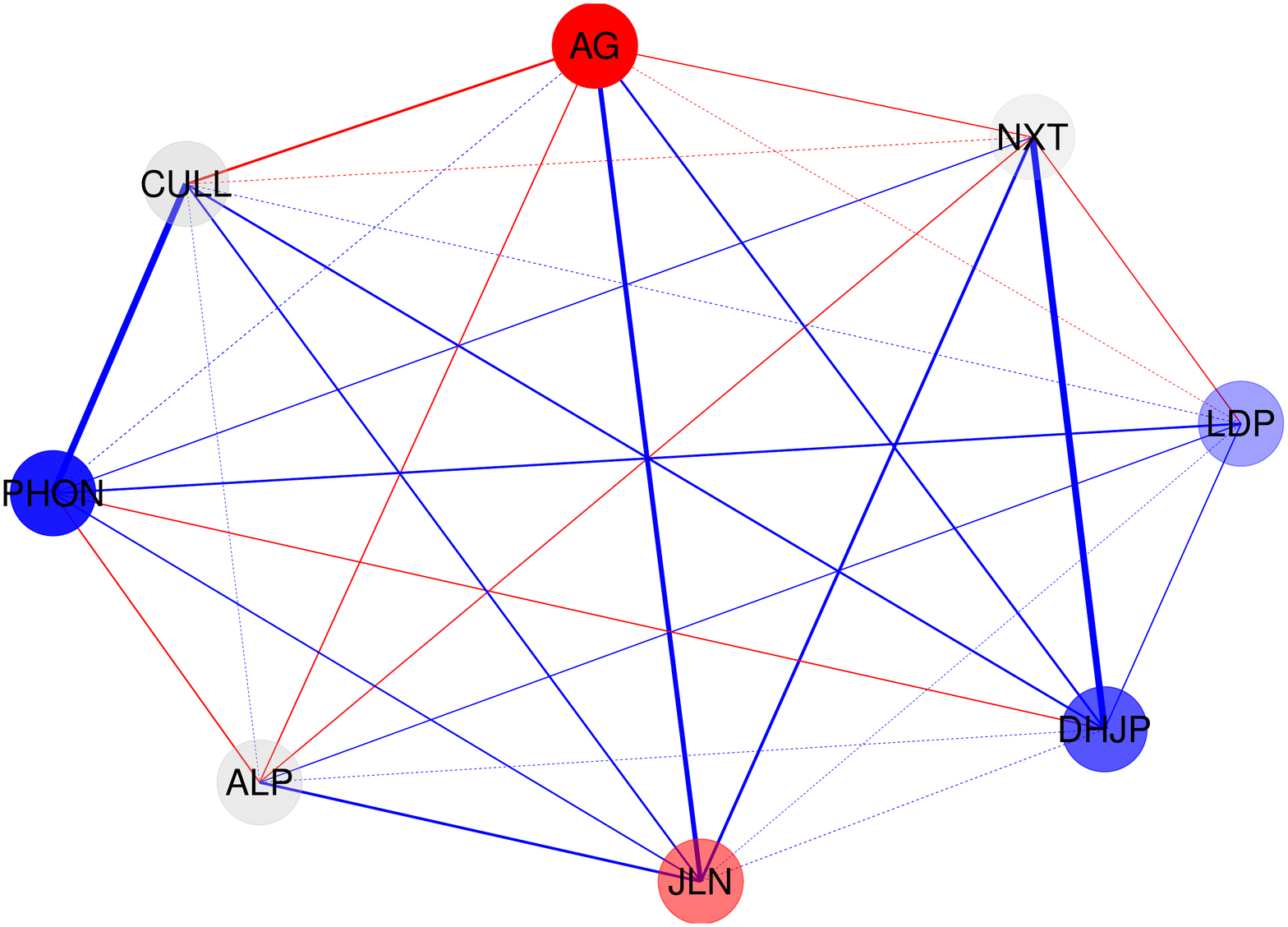}
\par\end{centering}
\caption{\label{fig: Network}Visualization of the significance and interaction
results from Tables \ref{tab: Estimates} and \ref{tab: p-values}.
Descriptions of the various elements of the figure appear within the
text.}

\end{figure}

From the results of Tables \ref{tab: Estimates} and \ref{tab: p-values},
we can construct the network diagram that is presented in Figure \ref{fig: Network}.
We can interpret the diagram as follows, with hypotheses declared
significant or otherwise at the $\alpha=0.05$ level. Each node is
colored blue, grey, or red, depending on whether the estimated bias
of the corresponding party is significant with a positive bias, insignificant,
or significant with a negative bias, respectively. Each node is more
opaque if the absolute value of the estimated bias is higher.

The edges connecting each node are either solid or dashed. A solid
edge indicates that the interaction between the two connected parties
is significant, and a dashed edge indicates that the interaction is
insignificant. An edge is colored blue if the interaction between
the two parties is positive, and a red edge indicates a negative interaction.
The edge thickness is directly proportional to the negative logarithm
of the $p\text{-value}$ of the corresponding interaction between
the two parties.

We note that 5 out of the 8 bias elements are inferred to be significant
different from zero, and 10 out of the 28 interaction elements are
inferred to be significantly different to zero. Given such a large
number of simultaneous hypothesis tests, and such a large number of
significant results, it is prudent to control for the potentially
inflated numbers of false positives that may occur. Towards this goal,
we choose to employ the paradigm of false discovery rate (FDR) control,
as considered by \citet{Benjamini1995}. Since there are potential
dependencies between the hypotheses that are tested, we choose to
use the method of \citet{Benjamini2001}, which is able to control
the FDR in the presence of dependencies between hypotheses. We adjust
the $p\text{-values}$ so that we may control the FDR via the method
of \citet{Benjamini2001} via the \emph{p.adjust} function in \emph{R}.
The adjusted $p\text{-values}$ can be interpreted as per the unadjusted
$p\text{-values}$, albeit with cutoffs indicating rejection of hypotheses
at some level of FDR control rather than the usual significance level,
instead. The FDR adjusted version of Table \ref{tab: p-values} is
displayed in Table \ref{tab: FDR control}.

\begin{table}
\caption{\label{tab: FDR control}Sub-table A contains the FDR-adjusted $p\text{-value}$
for the tests that each bias vector element is equal to zero. Sub-table
B contains the FDR-adjusted $p\text{-value}$ for the tests that each
interaction matrix element is equal to zero.}

\centering{}{\footnotesize{}}%
\begin{tabular}{ccccccccc}
{\footnotesize{}A} &  &  &  &  &  &  &  & \tabularnewline
\hline 
{\footnotesize{}Party} & {\footnotesize{}ALP} & {\footnotesize{}AG} & {\footnotesize{}NXT} & {\footnotesize{}PHON} & {\footnotesize{}LDP} & {\footnotesize{}JLN} & {\footnotesize{}DHJP} & {\footnotesize{}CULL}\tabularnewline
\hline 
{\footnotesize{}$\text{adj-}p$} & {\footnotesize{}1.88E-01} & {\footnotesize{}1.24E-05} & {\footnotesize{}8.52E-01} & {\footnotesize{}6.41E-03} & {\footnotesize{}8.45E-02} & {\footnotesize{}2.13E-02} & {\footnotesize{}2.13E-02} & {\footnotesize{}7.45E-01}\tabularnewline
\hline 
 &  &  &  &  &  &  &  & \tabularnewline
{\footnotesize{}B} &  &  &  &  &  &  &  & \tabularnewline
\cline{1-8} 
{\footnotesize{}Party} & {\footnotesize{}ALP} & {\footnotesize{}AG} & {\footnotesize{}NXT} & {\footnotesize{}PHON} & {\footnotesize{}LDP} & {\footnotesize{}JLN} & {\footnotesize{}DHJP} & \tabularnewline
\cline{1-8} 
{\footnotesize{}AG} & {\footnotesize{}9.84E-01} &  &  &  &  &  &  & \tabularnewline
{\footnotesize{}NXT} & {\footnotesize{}1.00E+00} & {\footnotesize{}1.00E+00} &  &  &  &  &  & \tabularnewline
{\footnotesize{}PHON} & {\footnotesize{}7.45E-01} & {\footnotesize{}1.00E+00} & {\footnotesize{}1.00E+00} &  &  &  &  & \tabularnewline
{\footnotesize{}LDP} & {\footnotesize{}1.00E+00} & {\footnotesize{}1.00E+00} & {\footnotesize{}1.00E+00} & {\footnotesize{}2.33E-01} &  &  &  & \tabularnewline
{\footnotesize{}JLN} & {\footnotesize{}1.29E-01} & {\footnotesize{}1.84E-03} & {\footnotesize{}7.96E-02} & {\footnotesize{}6.75E-01} & {\footnotesize{}1.00E+00} &  &  & \tabularnewline
{\footnotesize{}DHJP} & {\footnotesize{}1.00E+00} & {\footnotesize{}2.33E-01} & {\footnotesize{}4.17E-05} & {\footnotesize{}1.00E+00} & {\footnotesize{}1.00E+00} & {\footnotesize{}1.00E+00} &  & \tabularnewline
{\footnotesize{}CULL} & {\footnotesize{}1.00E+00} & {\footnotesize{}1.75E-01} & {\footnotesize{}1.00E+00} & {\footnotesize{}4.17E-05} & {\footnotesize{}1.00E+00} & {\footnotesize{}2.82E-01} & {\footnotesize{}2.33E-01} & \tabularnewline
\cline{1-8} 
\end{tabular}{\footnotesize \par}
\end{table}

We observe that the FDR-adjusted $p\text{-values}$ indicate that,
when controlling at the $5\%$ FDR, we would only reject 4 out of
the 8 hypotheses regarding the bias elements, and only 3 out of 28
hypotheses relating to the interaction matrix. Thus, the use of FDR
adjustment significantly reduces our power for detecting an interesting
interaction.

It is well-known that the method of \citet{Benjamini2001} is very
conservative, especially when there is a potentially large number
of hypotheses where the null hypotheses are false. That is, when there
are many false null hypotheses, the method of \citet{Benjamini2001}
controls the FDR at a level that is substantially smaller than the
specified cutoff level (i.e.~$5\%$, above). To counteract this effect,
it is not uncommon to utilize an FDR cutoff that is larger than the
significance level $\alpha$, that would otherwise be used in a significance
test. Thus, we choose to control the FDR at the $10\%$ level, which
yields 5 out of 8 rejected bias elements, and 4 out of 28 rejected
interaction elements.

\begin{figure}
\begin{centering}
\includegraphics[width=12cm]{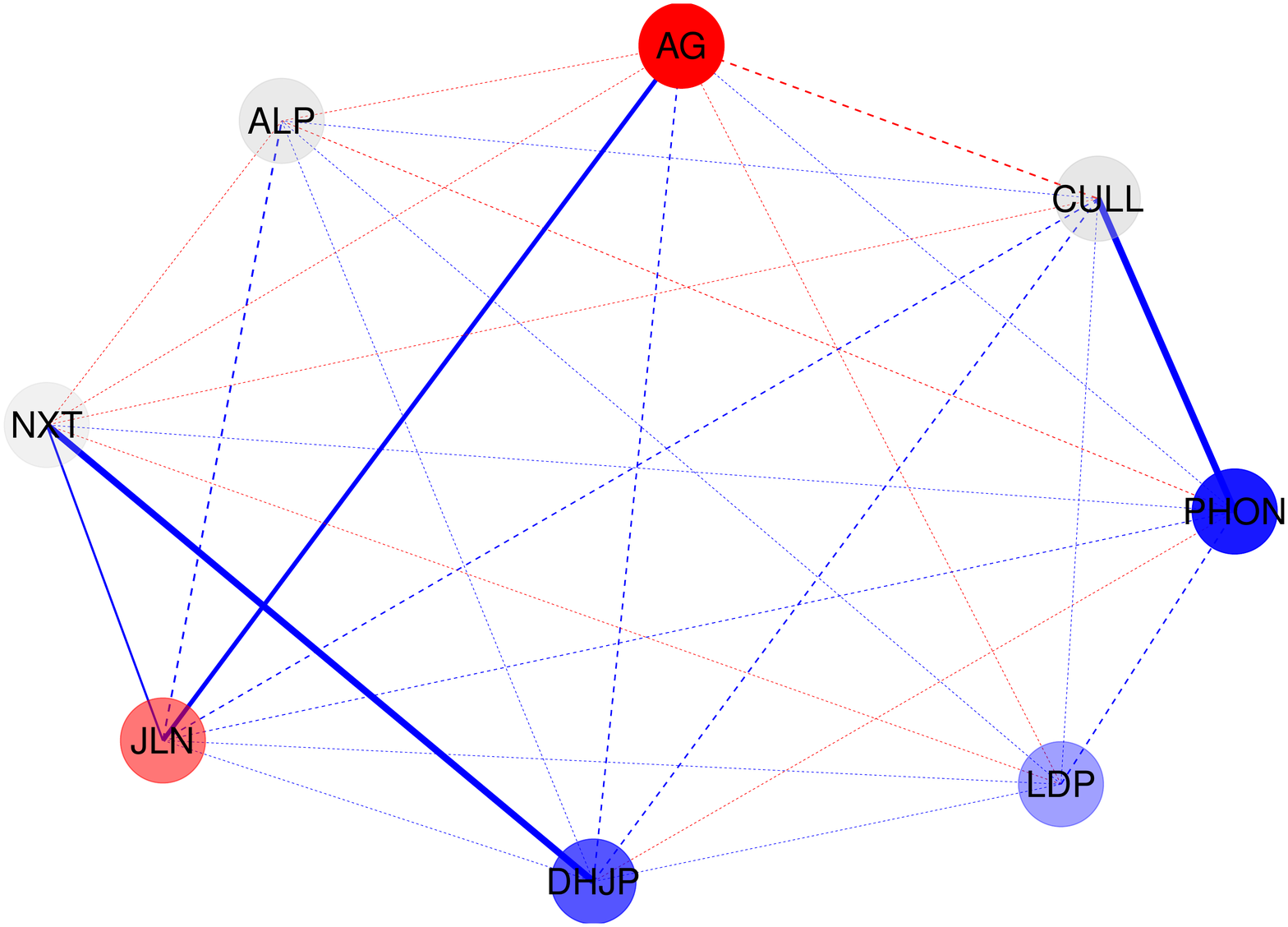}
\par\end{centering}
\caption{\label{fig: FDR Net}Visualization of the FDR control and interaction
results from Tables \ref{tab: Estimates} and \ref{tab: FDR control}.
Descriptions of the various elements of the figure appear within the
text.}
\end{figure}

As with Figure \ref{fig: Network}, we summarize the results of Tables
\ref{tab: Estimates} and \ref{tab: FDR control} in Figure \ref{fig: FDR Net}.
We can interpret the diagram as follows, with hypotheses declared
rejected or otherwise at the $10\%$ level of FDR control. Each node
is colored blue, grey, or red, depending on whether the hypothesis
of the estimated bias of the corresponding party is rejected with
a positive bias, not rejected, or rejected with a negative bias, respectively.
Each node is more opaque if the absolute value of the estimated bias
is higher.

The edges connecting each node are either solid or dashed. A solid
edge indicates that the corresponding hypothesis of the interaction
between the two connected parties is rejected, and a dashed edge indicates
that the hypothesis is not. An edge is colored blue if the interaction
between the two parties is positive, and a red edge indicates a negative
interaction. The edge thickness is directly proportional to the negative
logarithm of the adjusted $p\text{-value}$ of the corresponding interaction
between the two parties.

\subsection{Discussions}

We begin by discussing the results from Table \ref{tab: Probabilities}.
The ALP is the oldest of the Australian parties and has a continuous
existence since 1891, where it has either governed or acted as the
opposition \citep{Weller2003}. As opposition, it is unsurprising
that the ALP should have a low marginal probability of agreement with
the government. It is also unsurprising that AG has a low marginal
probability of agreement with the Government, since it is a progressive
party with ideologies that are antithetical to neo-liberalism and
conservatism of the governing LNP (see, e.g., \citealp{Miragliotta2010}
and \citealp{Mendes1998} regarding the AG and LNP, respectively).
Next, the high rates of agreement with the government, of PHON, LDP,
and CULL is also not surprising since PHON is largely a nationalistic
and conservative party \citep{Grant2019} and the LDP is self-described
as an economically neo-classically liberal and libertarian party (cf.~\url{www.ldp.org.au/our_philosophy}). 

The low probability of agreement between JLN and the Government is
somewhat surprising since JLN is seen as a conservative party (cf.~\citealp{Wood2018}).
Both NXT and DHJP are largely personality-based parties, with NXT
being self identified as being a centralist party that seeks to break
the duopoly of the ALP and LNP \citep{Kefford2018}. It is therefore
unsurprising that NXT has an agreement probability that is close to
$50\%$. On the other hand, DHJP ran a single-issue policy platforms
centered around judicial reform (cf.~\url{www.justiceparty.com.au/our-policies}),
and thus we had no a priori expectations regarding the relationship
between DHJP and the conservative Government. It is therefore of great
interest to observe that DHJP is generally amenable to the position
of the Government.

Next, we discuss the results from Tables \ref{tab: Estimates} and
\ref{tab: FDR control}, via the aid of Figure \ref{fig: FDR Net}.
We observe that upon controlling the FDR at the $10\%$ level and
in the presence of the interactions, PHON, LDP, and DHJP are biased
towards agreement with the Government, and AG and JLN are biased towards
disagreement with the Government. We also observe that ALP, NXT, and
CULL have no bias either way.

The biased positions of PHON and LDP towards agreement, and AG towards
disagreement is not surprising, as has been discussed above. However,
the bias against the Government by JLN is surprising, but is in correspondence
with what we observed from Table \ref{tab: Probabilities}. The bias
towards agreement by DHJP is also in correspondence with the above
$50\%$ marginal rate of agreement, that was observed in Table \ref{tab: Probabilities}.

Of the three parties that exhibit a lack of bias, when controlled
at the $10\%$ FDR level, the result for NXT and CULL are not surprising.
As we have already discussed, NXT and CULL are both parties that are
heavily driven by interactions with other parties (PHON and DHJP,
respectively). Given the strong bias towards voting for the government
of both PHON and DHJP, it is unsurprising that these strong interactions
drive the marginal rate of agreement of both NXT and CULL, upwards.

It is however surprising that ALP should lack a bias, given that they
are the opposition to the Government. We observe from Table \ref{tab: p-values}
that the downward bias from the ALP is significant at the $\alpha=0.1$
level. We thus propose that the lack of rejection of the hypothesis
for the bias element associated with the ALP may be due to sampling
error. Another potential reason is that the ALP and LNP have had a
history of voting together on issues such as defense, in the spirit
of bipartisanship (see, e.g., \citealp{Murphy2014} and \citealp{Carr2015}).
It is a matter of debate as to whether this bipartisanship represents
a lack of ideological difference between the two parties or whether
the behavior is merely pragmatic in nature.

We now discuss the interactions between the parties. The least surprising
of the interactions is that between PHON and CULL, given that they
voted the same way on all 12 Senate questions. Next, the positive
interaction between AG and JLN is perhaps due to the fact that Jacqui
Lambie, along with two of the nine AG Senators are Tasmanian (cf.~\citealp[Tab. 1]{Evans2016}
and \citealp{Bolwell2017}), and thus are allied in voting in a manner
that best benefits the state. 

The positive interaction between NXT and JLN is more obvious as both
Nick Xenophon as the two parties had agreed to an alliance prior to
the 2016 election (cf.~\citealp{Atkin2015}). Lastly, the positive
interaction between NXT and DHJP may come down to the two entities
having formed an alliance in the Senate (cf.~\citealp{Hinch2016}).

\section{Conclusion}

We had set out with the aim to quantitatively analyze the degree to
which each of the non-government parties of the Senate of the 45th
Australian Parliament were pro- or anti-Government, and the degree
to which the votes of each of the non-government Senate parties were
in concordance or discordance with one another. Using the FVBM as
an inferential vehicle, we were able to satisfactorily achieve both
of the stated aims.

Our investigation centered around the analysis of the Senate data,
that were described in Section \ref{sec: Data set}. We found that
the use of an FVBM, fitted via MPLE, provided a good fit to the data.
Furthermore, analysis of these data revealed voting patterns that
were largely supported by the literature and external sources. In
particular, our analyses of whether parties were pro- or anti-Government
yielded results that were in correspondence with the ideological positions
of the assessed parties. Furthermore, identified interactions between
parties were in general accordance with identifiable parliamentary
alliances. Thus, we conclude that our FVBM methodology was successfully
applied for the analysis of the Senate data, and may be applicable
to similarly structured data sets.

A number of possible future directions for our research have been
identified. Firstly, we may consider the case of dependent observations,
and allow for autoregressive structures between voting events that
follow sequentially from each other. Secondly, we may allow for the
bias and interaction parameter elements of the FVBM to be parametrizable
by covariates, in order to produce a richer class of models. Thirdly,
we may substitute the FDR control procedure with a regularization
procedure, instead, using penalizations such as the LASSO of \citet{Tibshirani1996}.
Such a procedure may be more powerful in identifying relationships
between political entities. Finally, given the parametric construction
of the FVBM, it is conceivable that we may be able to conduct missing
data imputation and parameter estimation, simultaneously. The derivation
of such an algorithm, as well as progress with respect to the other
future directions, would require significant further technical developments,
which cannot be achieved within the length and scope constraints of
this manuscript.

\section*{Appendix}

\subsection*{Sensitivity of the analysis to $k$ in $k\text{-nearest neighbor}$
imputation}

As discussed in Section \ref{subsec: Data-processing}, $k\text{-nearest neighbor}$
imputation was used to account for missing data in our Senate voting
data set, where $k=3$. The choice of $k=3$ was made based on a recommendation
by \citet{Beretta2016}, who observed that the choice provided a good
balance between precision and generalizability. 

In order to assess whether our analyses was sensitive to the choice,
we also conducted $k=1$ and $k=5$ imputation, and fitted FVBMs to
our differently imputed data. We reproduce Tables \ref{tab: Estimates}
and \ref{tab: FDR control} for the two alternative choices of $k$.
Tables \ref{tab: Estimates 1nn} and \ref{tab: FDR control 1nn} contain
the MPLE and FDR-adjusted $p\text{-values}$, computed after $1\text{-nearest neighbor}$
imputation. Tables \ref{tab: Estimates 5nn} and \ref{tab: FDR control 5nn}
contain the MPLE and FDR-adjusted $p\text{-values}$, computed after
$5\text{-nearest neighbor}$ imputation.

\begin{table}
\caption{\label{tab: Estimates 1nn}Sub-table A contains the estimated values
of the bias vector (i.e.~$\tilde{\bm{b}}_{n}$), computed after $1\text{-nearest neighbor}$
imputation. Sub-table B contains the corresponding estimated values
of the unique elements of the interaction matrix (i.e $\tilde{\bm{M}}_{n}$).}

\centering{}%
\begin{tabular}{ccccccccc}
A &  &  &  &  &  &  &  & \tabularnewline
\hline 
Party & ALP & AG & NXT & PHON & LDP & JLN & DHJP & CULL\tabularnewline
\hline 
Bias & -0.250 & -1.031 & -0.150 & 1.026 & 0.180 & -0.636 & 0.650 & -0.352\tabularnewline
\hline 
 &  &  &  &  &  &  &  & \tabularnewline
B &  &  &  &  &  &  &  & \tabularnewline
\cline{1-8} 
Party & ALP & AG & NXT & PHON & LDP & JLN & DHJP & \tabularnewline
\cline{1-8} 
AG & -0.306 &  &  &  &  &  &  & \tabularnewline
NXT & -0.209 & -0.272 &  &  &  &  &  & \tabularnewline
PHON & -0.345 & 0.065 & 0.327 &  &  &  &  & \tabularnewline
LDP & 0.093 & 0.062 & -0.103 & 0.654 &  &  &  & \tabularnewline
JLN & 0.498 & 0.752 & 0.405 & 0.506 & -0.069 &  &  & \tabularnewline
DHJP & 0.044 & 0.594 & 0.672 & -0.525 & 0.269 & 0.067 &  & \tabularnewline
CULL & 0.014 & -0.743 & -0.049 & 1.225 & 0.169 & 0.412 & 0.746 & \tabularnewline
\cline{1-8} 
\end{tabular}
\end{table}

\begin{table}
\caption{\label{tab: FDR control 1nn}Sub-table A contains the FDR-adjusted
$p\text{-value}$ for the tests that each bias vector element is equal
to zero, computed after $1\text{-nearest neighbor}$ imputation. Sub-table
B contains the corresponding FDR-adjusted $p\text{-value}$ for the
tests that each interaction matrix element is equal to zero.}

\centering{}{\footnotesize{}}%
\begin{tabular}{ccccccccc}
{\footnotesize{}A} &  &  &  &  &  &  &  & \tabularnewline
\hline 
{\footnotesize{}Party} & {\footnotesize{}ALP} & {\footnotesize{}AG} & {\footnotesize{}NXT} & {\footnotesize{}PHON} & {\footnotesize{}LDP} & {\footnotesize{}JLN} & {\footnotesize{}DHJP} & {\footnotesize{}CULL}\tabularnewline
\hline 
{\footnotesize{}$\text{adj-}p$} & {\footnotesize{}5.12E-01} & {\footnotesize{}1.83E-05} & {\footnotesize{}1.00E+00} & {\footnotesize{}2.69E-03} & {\footnotesize{}8.86E-01} & {\footnotesize{}6.92E-03} & {\footnotesize{}2.60E-02} & {\footnotesize{}8.45E-01}\tabularnewline
\hline 
 &  &  &  &  &  &  &  & \tabularnewline
{\footnotesize{}B} &  &  &  &  &  &  &  & \tabularnewline
\cline{1-8} 
{\footnotesize{}Party} & {\footnotesize{}ALP} & {\footnotesize{}AG} & {\footnotesize{}NXT} & {\footnotesize{}PHON} & {\footnotesize{}LDP} & {\footnotesize{}JLN} & {\footnotesize{}DHJP} & \tabularnewline
\cline{1-8} 
{\footnotesize{}AG} & {\footnotesize{}2.87E-01} &  &  &  &  &  &  & \tabularnewline
{\footnotesize{}NXT} & {\footnotesize{}6.98E-01} & {\footnotesize{}1.00E+00} &  &  &  &  &  & \tabularnewline
{\footnotesize{}PHON} & {\footnotesize{}6.98E-01} & {\footnotesize{}1.00E+00} & {\footnotesize{}1.00E+00} &  &  &  &  & \tabularnewline
{\footnotesize{}LDP} & {\footnotesize{}1.00E+00} & {\footnotesize{}1.00E+00} & {\footnotesize{}1.00E+00} & {\footnotesize{}5.00E-02} &  &  &  & \tabularnewline
{\footnotesize{}JLN} & {\footnotesize{}1.73E-03} & {\footnotesize{}6.92E-05} & {\footnotesize{}1.54E-01} & {\footnotesize{}2.87E-01} & {\footnotesize{}1.00E+00} &  &  & \tabularnewline
{\footnotesize{}DHJP} & {\footnotesize{}1.00E+00} & {\footnotesize{}1.93E-01} & {\footnotesize{}9.47E-05} & {\footnotesize{}6.98E-01} & {\footnotesize{}4.86E-01} & {\footnotesize{}1.00E+00} &  & \tabularnewline
{\footnotesize{}CULL} & {\footnotesize{}1.00E+00} & {\footnotesize{}1.54E-01} & {\footnotesize{}1.00E+00} & {\footnotesize{}6.92E-05} & {\footnotesize{}1.00E+00} & {\footnotesize{}2.87E-01} & {\footnotesize{}1.93E-01} & \tabularnewline
\cline{1-8} 
\end{tabular}{\footnotesize \par}
\end{table}

\begin{table}
\caption{\label{tab: Estimates 5nn}Sub-table A contains the estimated values
of the bias vector (i.e.~$\tilde{\bm{b}}_{n}$), computed after $5\text{-nearest neighbor}$
imputation. Sub-table B contains the corresponding estimated values
of the unique elements of the interaction matrix (i.e $\tilde{\bm{M}}_{n}$).}

\centering{}%
\begin{tabular}{ccccccccc}
A &  &  &  &  &  &  &  & \tabularnewline
\hline 
Party & ALP & AG & NXT & PHON & LDP & JLN & DHJP & CULL\tabularnewline
\hline 
Bias & -0.326 & -1.077 & -0.185 & 0.830 & 0.532 & -0.555 & 0.751 & -0.433\tabularnewline
\hline 
 &  &  &  &  &  &  &  & \tabularnewline
B &  &  &  &  &  &  &  & \tabularnewline
\cline{1-8} 
Party & ALP & AG & NXT & PHON & LDP & JLN & DHJP & \tabularnewline
\cline{1-8} 
AG & -0.209 &  &  &  &  &  &  & \tabularnewline
NXT & -0.197 & -0.256 &  &  &  &  &  & \tabularnewline
PHON & -0.355 & 0.058 & 0.349 &  &  &  &  & \tabularnewline
LDP & 0.149 & 0.121 & -0.199 & 0.532 &  &  &  & \tabularnewline
JLN & 0.312 & 0.623 & 0.421 & 0.358 & 0.033 &  &  & \tabularnewline
DHJP & 0.068 & 0.593 & 0.791 & -0.488 & 0.050 & 0.093 &  & \tabularnewline
CULL & 0.053 & -0.754 & -0.119 & 1.226 & 0.314 & 0.395 & 0.824 & \tabularnewline
\cline{1-8} 
\end{tabular}
\end{table}

\begin{table}
\caption{\label{tab: FDR control 5nn}Sub-table A contains the FDR-adjusted
$p\text{-value}$ for the tests that each bias vector element is equal
to zero, computed after $5\text{-nearest neighbor}$ imputation. Sub-table
B contains the corresponding FDR-adjusted $p\text{-value}$ for the
tests that each interaction matrix element is equal to zero.}

\centering{}{\footnotesize{}}%
\begin{tabular}{ccccccccc}
{\footnotesize{}A} &  &  &  &  &  &  &  & \tabularnewline
\hline 
{\footnotesize{}Party} & {\footnotesize{}ALP} & {\footnotesize{}AG} & {\footnotesize{}NXT} & {\footnotesize{}PHON} & {\footnotesize{}LDP} & {\footnotesize{}JLN} & {\footnotesize{}DHJP} & {\footnotesize{}CULL}\tabularnewline
\hline 
{\footnotesize{}$\text{adj-}p$} & {\footnotesize{}1.88E-01} & {\footnotesize{}2.45E-05} & {\footnotesize{}1.00E+00} & {\footnotesize{}3.45E-02} & {\footnotesize{}3.45E-02} & {\footnotesize{}2.82E-02} & {\footnotesize{}2.82E-02} & {\footnotesize{}5.90E-01}\tabularnewline
\hline 
 &  &  &  &  &  &  &  & \tabularnewline
{\footnotesize{}B} &  &  &  &  &  &  &  & \tabularnewline
\cline{1-8} 
{\footnotesize{}Party} & {\footnotesize{}ALP} & {\footnotesize{}AG} & {\footnotesize{}NXT} & {\footnotesize{}PHON} & {\footnotesize{}LDP} & {\footnotesize{}JLN} & {\footnotesize{}DHJP} & \tabularnewline
\cline{1-8} 
{\footnotesize{}AG} & {\footnotesize{}9.20E-01} &  &  &  &  &  &  & \tabularnewline
{\footnotesize{}NXT} & {\footnotesize{}9.97E-01} & {\footnotesize{}1.00E+00} &  &  &  &  &  & \tabularnewline
{\footnotesize{}PHON} & {\footnotesize{}9.13E-01} & {\footnotesize{}1.00E+00} & {\footnotesize{}1.00E+00} &  &  &  &  & \tabularnewline
{\footnotesize{}LDP} & {\footnotesize{}1.00E+00} & {\footnotesize{}1.00E+00} & {\footnotesize{}1.00E+00} & {\footnotesize{}2.27E-01} &  &  &  & \tabularnewline
{\footnotesize{}JLN} & {\footnotesize{}1.48E-01} & {\footnotesize{}2.05E-03} & {\footnotesize{}7.37E-02} & {\footnotesize{}9.13E-01} & {\footnotesize{}1.00E+00} &  &  & \tabularnewline
{\footnotesize{}DHJP} & {\footnotesize{}1.00E+00} & {\footnotesize{}2.27E-01} & {\footnotesize{}2.97E-05} & {\footnotesize{}1.00E+00} & {\footnotesize{}1.00E+00} & {\footnotesize{}1.00E+00} &  & \tabularnewline
{\footnotesize{}CULL} & {\footnotesize{}1.00E+00} & {\footnotesize{}1.64E-01} & {\footnotesize{}1.00E+00} & {\footnotesize{}2.07E-04} & {\footnotesize{}1.00E+00} & {\footnotesize{}2.98E-01} & {\footnotesize{}2.27E-01} & \tabularnewline
\cline{1-8} 
\end{tabular}{\footnotesize \par}
\end{table}

From Tables \ref{tab: Estimates 1nn} and \ref{tab: Estimates 5nn},
we observe that, although there are small deviations in the value
of the MPLE estimates, the direction of the estimated biases and interactions
are the same as those that are reported in Table \ref{tab: Estimates}.
The results from Table \ref{tab: FDR control 5nn} show that the conclusions
drawn with FDR controlled at the $10\%$ level, under $5\text{-nearest neighbor}$
imputation, is the same as the conclusions that can be drawn from
Table \ref{tab: FDR control}.

We observe that there are some small differences in conclusions, when
drawing inference with FDR controlled at the $10\%$ level, between
Tables \ref{tab: FDR control} and \ref{tab: FDR control 1nn}. Firstly,
we no longer reject the assumption of no bias for LDP, and we no longer
infer interactions between JLN and NXT. We reject an additional two
interaction hypotheses, however, between ALP and JLN, and between
PHON and LDP. None of these changes contradict the conclusions that
we have drawn in Section \ref{sec: Analysis}.

Simulation results from \citet{Beretta2016} show that $1\text{-nearest neighbor}$
imputation can often be inaccurate, when compared to larger values
of $k$. However, the simulations of \citet{Beretta2016} were for
a real-valued regression problem, rather than a binary PMF estimation
problem. Thus, we are cautious to make any stronger claims regarding
the relevance of their results to our modeling problem. We believe
that a model-based approach that conducts imputation and estimation,
simultaneously, may resolve our problem of having to choose tune our
imputation scheme. However, as we have already mentioned in our conclusion,
this would constitute a significant amount of new work that lies outside
the scope of this manuscript.

\section*{Acknowledgments}

Jessica Bagnall is funded by a Research Training Program Stipend scholarship
from La Trobe University. Hien Nguyen is personally funded by Australian
Research Council (ARC) DECRA fellowship DE170101134 and a La Trobe
University startup grant.

\bibliographystyle{plainnat}
\bibliography{FVBM}

\end{document}